\documentclass{rmf-d}
\usepackage{nopageno,rmfbib,multicol,times,epsf,amsmath,amssymb,cite}
\usepackage[]{caption2}
\usepackage{color}
\usepackage[dvipsnames]{xcolor}

\usepackage{appendix}
\usepackage{soul}
\usepackage{ulem}
\usepackage{graphicx}
\usepackage{graphics}
\usepackage{tikz}
\usepackage{longtable}
\usepackage{comment}
\usepackage{hyperref}
\usepackage{xfrac}
\usepackage{multirow}% http://ctan.org/pkg/multirow
\usepackage{hhline}% http://ctan.org/pkg/hhline
\usepackage{makecell}
\newcommand{\mc}[3]{\multicolumn{#1}{#2}{#3}}
\def\rmfcornisa{Research OR Education of Physics \hfill\rmf\ {\bf ??} (*?*) ???--???
\hfill MES? A\~NO?}

\def\rmfcintilla{{\it Rev.\ Mex.\ Fis.\/} {\bf ??} (*?*) (????) ???--???}
\clearpage \rmfcaptionstyle 
\begin{document}
\title{   Double parton scattering via photon-proton interactions: a new light
on the transverse proton structure
\vspace{-6pt}}
\author{ Matteo Rinaldi     }
\address{ Dipartimento di Fisica e Geologia. Universit\`a degli studi di 
Perugia. INFN section of Perugia. Via A. Pascoli, Perugia, Italy.    }
\author{ }
\address{ }
\author{ }
\address{ }
\author{ }
\address{ }
\author{ }
\address{ }
\author{ }
\address{ }
\maketitle
\recibido{day month year}{day month year
\vspace{-12pt}}
\begin{abstract}
\vspace{1em}

In this contribution
we
present the main results of the investigation about
 double parton scattering (DPS) in
quasi-real photon-proton interactions.
We show the first evaluation of the DPS cross-section at leading-order
for the four-jet photo-production observed at HERA.
To this aim the $\gamma-p$  effective cross section has been computed
for the first time.
 One of the main outcomes of this
analysis is that the DPS contribution is not negligible and potentially
measurable. Furthermore,  possible future data could be used to 
get new information on
the transverse proton structure  not accessible in other
processes.

  \vspace{1em}
\end{abstract}
\keys{  PLEASE PROVIDE SUMMARY IN ENGLISH  \vspace{-4pt}}
\pacs{   \bf{\textit{PLEASE PROVIDE }}    \vspace{-4pt}}
\begin{multicols}{2}

%%%%%%%%%%%%%%%%%%%%%%% INIZIO

\section{Introduction}
Here we discuss  the main outcomes of Refs. \cite{Rinaldi:2021vbj}.
In collision involving hadrons, the role of
 multiple parton interactions
 (MPI), due to extending nature hadrons,
 has been established \cite{Goebel:1979mi,
Humpert:1983pw,Mekhfi:1983az,Mekhfi:1985dv,Humpert:1984ay,
Mangano:1988sq,1a,Sjostrand:1986ep}.
Here we focus on
 double parton scattering (DPS). In fact, new non-perturbative
information on the structure of proton, not accessible in
single parton scattering (SPS),
 can be obtained \cite{rapid,jhepc,noij2,noij1}.
Indeed, the DPS cross-section depends on the almost unknown
double Parton Distribution Functions
(dPDFs), i.e.,
the  number densities of a parton pair  with a given  transverse distance
$b_\perp$  and carrying  longitudinal momentum fractions
($ x_1,x_2$) of the parent hadron \cite{4a,Diehl:2011tt,5a,noi1,blok_1,Blok1
,Blok2,gauntevo,man,noij2,noij1,Diehl:2020xyg,Diehl:2018kgr,Diehl:2017kgu,
Diehl:2015bca,Gaunt:2014ska,Diehl:2019rdh,Diehl:2018wfy,
Ryskin:2011kk,Ryskin:2012qx}. Up today, data on dPDFs are not available and 
usually, the experimental findings are collected in the
so-called
 $\sigma^{pp}_{eff}$,
 in $pp$ collisions \cite{Calucci:1999yz},
 { and recently in $pA$ collisions \cite{Aaij:2020smi}}.
This quantity controls the magnitude of DPS contribution under
 the assumptions of two uncorrelated hard scatterings and full
 factorisation of dPDFs in terms of ordinary PDFs.
It has been shown that the knowledge of
 $\sigma^{pp}_{eff}$  can provide
information on the proton structure  \cite{rapid,jhepc}. We remind that from data
$\sigma_{eff} \sim 8-35$ mb
 \cite{4jet,CMS:2021ijt,Chapon:2020heu}
In Ref. \cite{Rinaldi:2021vbj} we calculated the DPS
cross-section in $ep$ collisions and thus
photon-proton interactions. In fact,
 the splitting quasi-real photon can initiate the DPS \cite{Klasen:2002xb}.
 In particular we considered the
four-jet photoproduction observed at HERA and analysed
by the ZEUS collaboration \cite{Butterworth:1996zw}.
In this case, the DPS involves a photon of variable and controllable transverse size, depending on its virtuality $Q^2$.
Since a complete formulation of photon and proton dPDFs  is missing
 at the moment,  we elaborate on a much
simpler quantity, $\sigma^{\gamma p}_{eff}$. We also include in the analysis
the estimate of main background, i.e.,  the SPS four-jet
photo-production cross-section \cite{Rinaldi:2021vbj}.
Furthermore,
we then show  that  the $Q^2$ dependence of $\sigma_{eff}^{\gamma p}$
is crucial to obtain
{ the first estimate of the mean transverse distance between partons in
the proton. }
To this aim we provided the
 necessary { integrated} luminosity to observe the  $Q^2$
dependence of the
DPS cross sections  in HERA kinematics.

\section{Effective cross-section for $\gamma p$ DPS }
\begin{figure*}[t]
\includegraphics[scale=0.6]{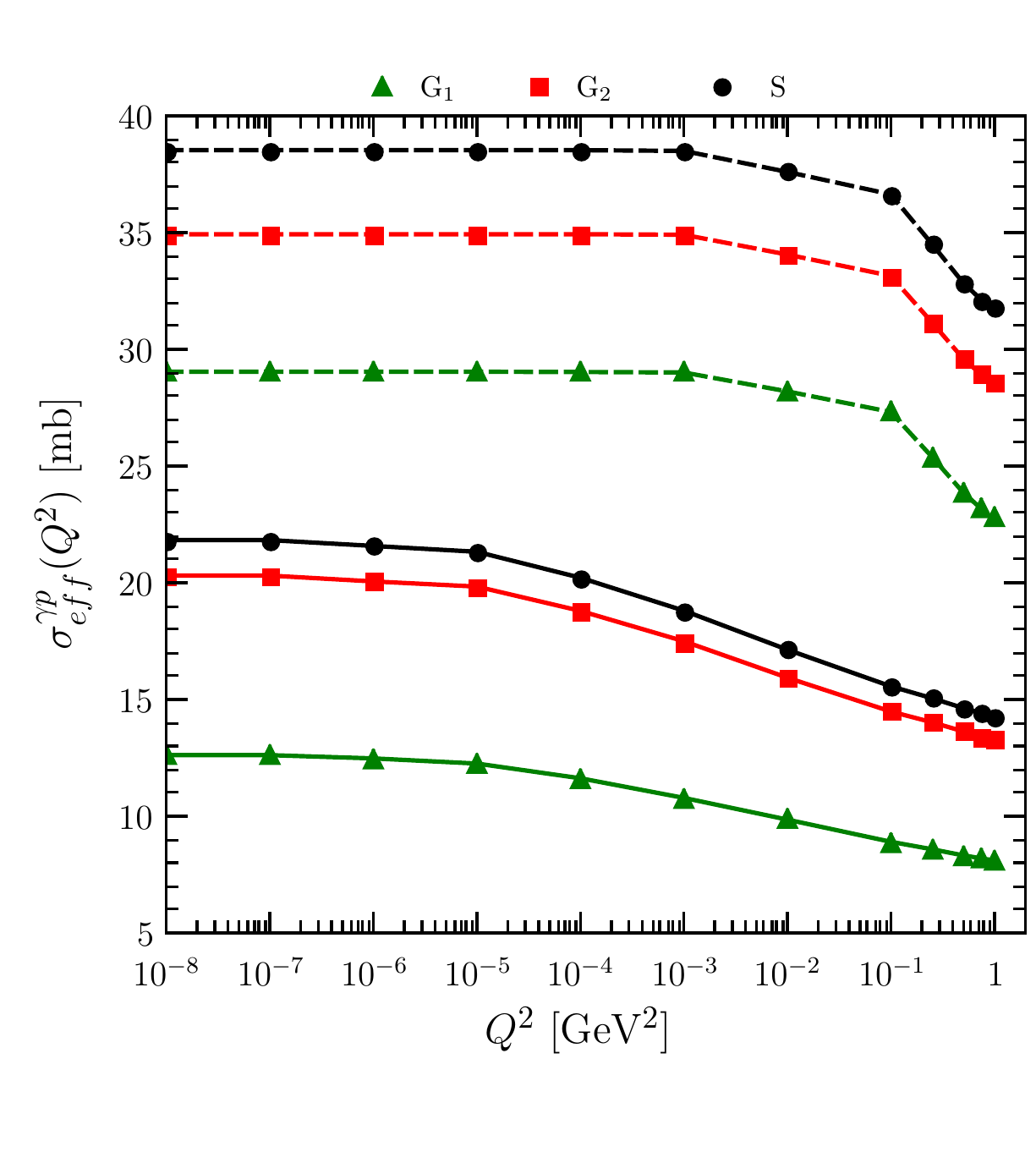} \hskip 0.5cm
 \includegraphics[scale=0.6]{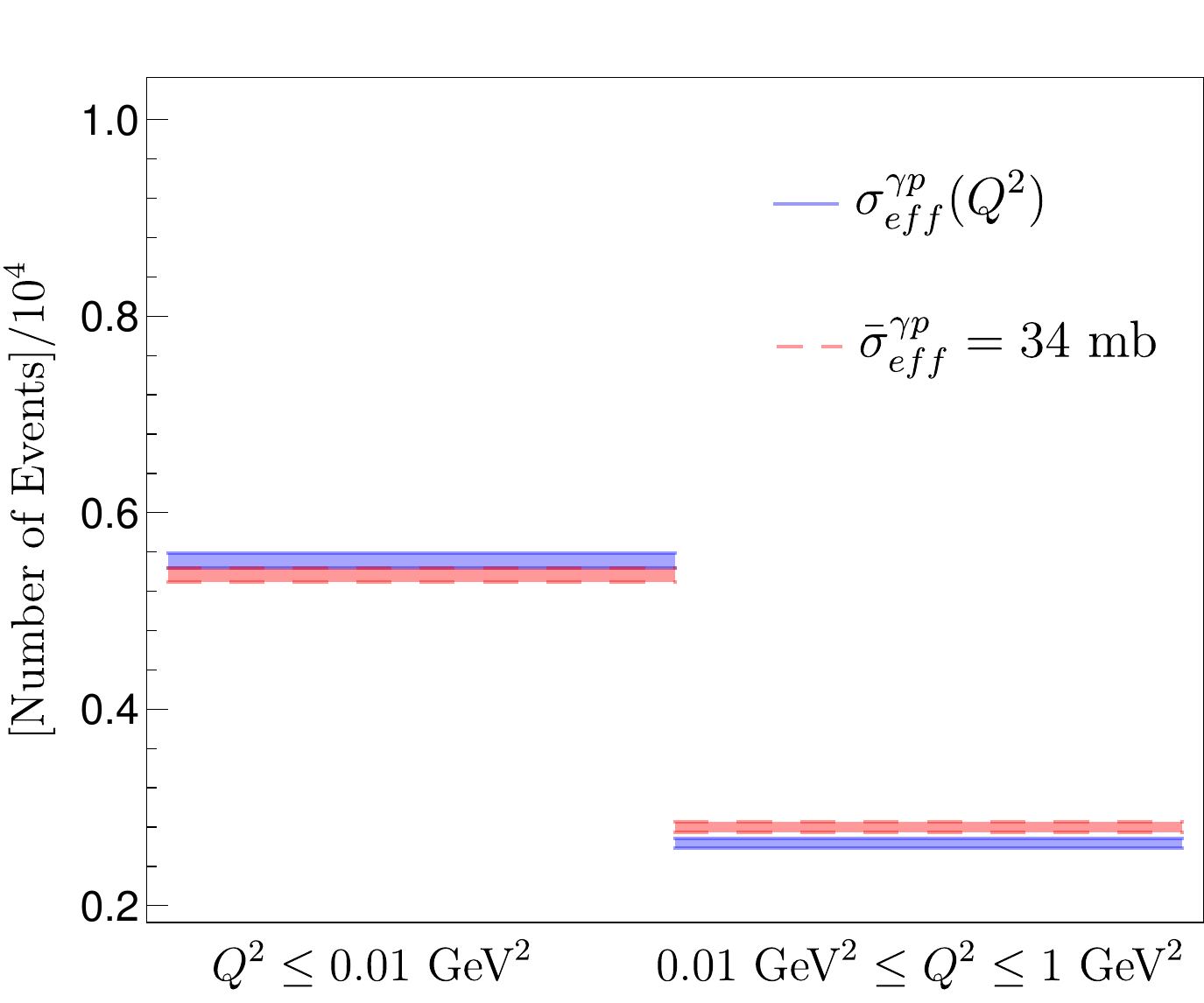}

\caption{\footnotesize $\sigma_{eff}^{\gamma p}$   evaluated in Eq.
(\ref{sigmaeff1})
 with the w.f. of Ref. \cite{arriola} (dashed lines) and  Ref.
\cite{brodfrank} (full lines) as a function
of $Q^2$. Different symbols denote the
proton effs described in the text.
The estimated number of events as a
function  of $Q^2$ for 200 $pb^{-1}$ of integrated luminosity for the photon
model of Ref. \cite{arriola} and proton eff G2.
 Full lines stand the  evaluations of $\sigma_{DPS}$ by means of
$\sigma_{eff}^{\gamma p}(Q^2)$ and
the
dotted ones represent the calculations of $\sigma_{DPS}$ by using the
$Q^2$-independent $\bar \sigma_{eff}^{\gamma p}$.}
%The transverse momentum cut-off is set to $k_{cut}=10^3$ GeV.
\label{Newfig}
\end{figure*}

%\section{$\sigma_{eff}^{\gamma p}$}
Here we introduce  $\sigma_{eff}^{\gamma p}$. One can
generalize the definition of the same quantity for $pp$ collisions \cite{rapid} which involves
the proton effective form factor (eff) \cite{noiPLB}. Starting from
the proton and photon dPDFs, i.e.,
{${D_{q_i q_j/p}}(x_i,x_j,k_\perp)$} and
${D_{q\bar q/\gamma}}(x_k,x_l,k_\perp)$, respectively where $ij$ and $kl$
are the flavours of the interacting partons, $k_\perp$
is the momentum imbalance, Fourier conjugate variable to the partonic transverse
distance, $b_\perp$, and $x$'s are the longitudinal momentum fractions carried
by each parton. For both mesons and baryons
\cite{noipion,Kasemets:2016nio,Rinaldi:2020ybv,noij1},  dPDFs can be calculated from
 the Light-Front (LF) wave functions for some quark
model. Then, one can define the so called
eff \cite{noiPLB}
which reads, e.g., for the photon:

\begin{align}
    F_2^{\gamma} (\vec k_\perp) =
  \frac{ \displaystyle  \sum_{ q}  \int ~dx~ {D_{q\bar q/\gamma}}
(x,\vec k_\perp;Q^2) }{\displaystyle  \sum_{q} \int ~dx~ {D_{q\bar q/\gamma}}
(x,\vec k_\perp=0;Q^2) }~.
\end{align}
This quantity is used to define $\sigma_{eff}^{\gamma p}$ in
the approximation
that momentum correlations and parton flavor dependence are neglected. Moreover,
if the proton dPDFs can be factorized in terms of PDFs,
one gets:
\begin{align}
    \label{sigmaeff1}
    \sigma_{eff}^{\gamma p}(Q^2) =  \left[
\int \dfrac{d^2k_\perp}{(2 \pi)^2} F_2^p(k_\perp) F_2^\gamma (k_\perp;Q^2)
\right]^{-1}\,.
\end{align}
In this scenario,
the $\gamma p$ DPS cross section for the production of
the final state $A+B$
is rearranged  in a pocket
formula $\sigma_{DPS}^{A+B} \sim \sigma_{SPS}^A
\sigma_{SPS}^B/\sigma_{eff}^{\gamma p}$.
Since the are no data on $\sigma_{eff}^{\gamma p}(Q^2)$, 
in Ref. \cite{Rinaldi:2021vbj} we calculated it by using the models of 
Refs. \cite{arriola,brodfrank} to describe both the photon
splitting mechanism. For the proton eff, we used
the approach of Ref. \cite{Blok2}, which we address as model
``S'', returning a $\sigma_{eff}^{pp}\sim 30$ mb.
In addition,
we also used a Gaussian ansatz.  The width of the
of this quantity, is a free parameter
 which produces  $\sigma_{eff}^{pp}=15$ mb
(``G$_1$'' model)
and    $\sigma_{eff}^{pp}=25$ mb
(``G$_2$'' model).
We remind that in the case of  Ref. \cite{brodfrank}, where the LO QED is used, 
a detail analysis on the regularization
procedure is provided in Ref. \cite{Rinaldi:2021vbj}
We present our numerical estimates
for $\sigma_{eff}^{\gamma p}(Q^2)$
in Fig. (\ref{Newfig}).
One may notice that the hadronic models
of Ref. \cite{arriola} systematically returns a
 higher $\sigma_{eff}^{\gamma p}$
with respect to that of Ref.
\cite{brodfrank}. Moreover, there is large sensitivity to the
proton eff.
We observe that, in the limit of high photon
virtuality, the value of $\sigma_{eff}^{\gamma p}$ can be predicted in complete
analogy with the gluon splitting case elaborated in Ref. \cite{Gaunt:2012dd},
see Ref. \cite{Rinaldi:2021vbj} for details.

\section{The geometry of
 $\sigma_{eff}^{\gamma p}(Q^2)$ }
\label{III}
Here we show how the quantity, previously presented, if measured, could unveil 
 new information on the proton structure. In fact
in hadron-hadron collisions, such a goal is almost prevented due to the lack of 
data on the proton eff \cite{rapid,jhepc}.
Nevertheless, in this case,
 ${\langle b^2_\perp \rangle_p}$, i.e. the main transverse distance
 between two
partons in the proton can be extracted from data on  $\sigma_{eff}^{\gamma p}$.
We consider
 $\tilde F_2(b_\perp)$,
  probability distribution of finding two partons at a given
transverse distance $b_\perp$
\cite{Calucci:1999yz,rapid,jhepc}, i.e., the Fourier Transform of the eff:

\begin{align}
    \label{rel1}
   \Big[ \sigma_{eff}^{\gamma p}(Q^2) \Big]^{-1} &=
 \int d^2 b_\perp~ \tilde F^p_2(b_\perp) \tilde F^{\gamma}_2(b_\perp;Q^2)
   \\
   \nonumber
   &= \sum_n C_n(\bar b_\perp;Q^2) \langle 
(b_\perp-\bar b_\perp)^n \rangle_p.
\end{align}
where, in the last passage, we Taylor expanded the photon distribution
around $\bar b_\perp$.
A realistic description of $C_n(\bar b_\perp;Q^2)$, together with  data on the
$Q^2$
dependence of $ \sigma_{eff}^{\gamma p}(Q^2) $, will allow to
 access  the transverse distance of partons in the proton.
In fact, {for a given specific dependence of $C_n$ on $Q^2$, one can identify
an}
 operator, $\mathcal{O}_{Q^2}^m$, such that
 % In fact,
%an operator could be identified so that:
\begin{align}
\label{ope}
   \mathcal{O}_{Q^2}^m  \Big[ \sigma_{eff}^{\gamma p}(Q^2) \Big]^{-1} =
 \mathcal{O}_{Q^2}^m C_m(\bar b_\perp,Q^2)  \langle (b_\perp-\bar b_\perp)^m
\rangle_p\,,
\end{align}
{and then } one {can select}  and extract
 $ \langle (b_\perp-\tilde b_\perp)^m
\rangle_p$, i.e. the
relevant information on the proton structure.
Details and examples of the application of this procedure are provided in Ref.
\cite{Rinaldi:2021vbj} since data on $\sigma_{eff}^{\gamma p}$ are not yet 
available.
The only practical limitation is represented by the accuracy with which
the dependence of $\sigma_{eff}^{\gamma p}$ on $Q^2$ could be 
eventually measured.
This relation strongly 
motivate this type of  measurements at facilities where the photon
virtuality can be experimentally measured such as the future  Electron Ion
Collider \cite{AbdulKhalek:2021gbh}. We show here an instructive example.

We consider a Gaussian photon effective form factor and thus the relative
normalized probability distribution reads:
\begin{align}
    \tilde F_2 ^\gamma(b;Q^2)=\dfrac{Q^2}{\alpha^2 \pi} 
e^{-b^2 Q^2/\alpha^2} ~.
\end{align}
The width depends on a free parameter $\alpha$.
The main transverse distance between the partons,
produced by the splitting mechanism, is:
\begin{align}
   \langle b^2 \rangle_\gamma= \int d^2b ~b^2 \tilde F_{2}^\gamma(b;Q^2)=
\dfrac{\alpha^2}{Q^2}~.
\end{align}
{One should notice that, despite the simplicity of the model,
 the mean distance between the
two produced partons goes to zero, as expected for high virtualities.  }
We can now  expand $\tilde F_2^ \gamma(b;Q^2)$:
\begin{align}
    \tilde F_{2}^ \gamma(b;Q^2) \sim  \dfrac{Q^2}{\pi \alpha^2}-\dfrac{Q^4}{\pi
\alpha^4} b^2+\mathcal{O}(b^4),
\end{align}
where $C_0(Q^2)=
\dfrac{Q^2}{\pi \alpha^2}$ and $C_2(Q^2) =- \dfrac{Q^4}{\pi \alpha^4}$. In this
scenario,
  a suitable operator which isolates
$\langle b^2 \rangle_p$ is $\mathcal{\hat
O}= d/(Q^3 dQ)|_{Q^2=0}$. With this choice, we can prove that:
\begin{align}
    &\frac{d}{Q^3 d Q} \Big( [\sigma_{eff}^{\gamma p}(Q^2)]^{-1}-C_0(Q^2) \Big)
\Big|_{Q^2=0} =
\\
\nonumber &  \frac{d}{Q^3 d Q} \Big( C_2(Q^2) \Big)  \Big|_{Q^2=0}
\langle b^2 \rangle_p~.
\end{align}
We also consider, for simplicity, that the proton distribution is:
\begin{align}
\tilde F_2^p(b)= e^{ -b^2 \beta^2  }\dfrac{\beta^2}{\pi}  \,,
\end{align}
and the mean transverse distance reads
    \begin{align}
          \int d^2b~  b^2 \tilde F_{2}(b)= \dfrac{1}{\beta^2}~.
    \end{align}
The 
$\gamma p$ effective cross section would be:
\begin{align}
   [\sigma_{eff}^{\gamma p}(Q^2)]^{-1} =  \dfrac{{\beta}^2 Q^2}{\pi
\left(\alpha^2
\beta^2+Q^2\right)}\,,
\end{align}
and therefore, the application of the operator leads to:
\begin{align}
     \frac{d}{Q^3 d Q} \Big( [\sigma_{eff}^{\gamma p}(Q^2)]^{-1}-C_0(Q^2) \Big)
\Big|_{Q^2=0} &= -\frac{4}{\pi  {\alpha}^4 {\beta}^2} \,,
     \\
      \frac{d}{Q^3 d Q} \Big( C_2(Q^2) \Big)  \Big|_{Q^2=0}
&=-\frac{4}{\pi  {\alpha}^4 } \,,
\end{align}
which can be combined to give
\begin{align}
    \langle b^2 \rangle_p =  \dfrac{\frac{d}{Q^3 d Q} \Big(
[\sigma_{eff}^{\gamma p}(Q^2)]^{-1}-C_0(Q^2) \Big) \Big|_{Q^2=0} }{\frac{d}{Q^3
d Q}
\Big( C_2(Q^2) \Big)  \Big|_{Q^2=0}} = \dfrac{1}{\beta^2}\,,
\end{align}
This is an example of how from the $Q^2$ dependence of $\sigma_{eff}^{\gamma 
p}$ one could extract proton information if the photon splitting mechanism is 
well established.

\section{The four-jet photo-production cross-section}

The four-jet  photoproduction at HERA has been investigated by the
ZEUS collaboration \cite{Chekanov:2007ab} by considering
jets with transverse energy
$E_T^{jet} > 6$ GeV and laboratory pseudorapidity
$|\eta_{jet} | < 2.4$, in the kinematic region  $Q^2 <
1 \; \mbox{GeV}^2$ and $y=E_\gamma/E_l$, i.e.
the energy fraction transferred between  the photon and lepton,
 in the range $0.2 \leq y \leq 0.85$. The collaboration pointed out how 
the inclusion of MPI
significantly improve the description of the data.
\cite{Marchesini:1991ch,Butterworth:1996zw,Sjostrand:2000wi}. Therefore, in 
Ref. \cite{Rinaldi:2021vbj},  we adopted the same
 kinematical cuts of Ref.
\cite{Chekanov:2007ab} together with the  pocket formula to evaluate 
 $\sigma_{DPS}$ is now \cite{Gaunt:2012dd}:

\begin{align}
\label{sigmaDPS}
d\sigma_{DPS}^{4j} = & \frac{1}{2}
\sum_{ab,cd} \int dy \; dQ^2 \;
\frac{f_{\gamma/e}(y,Q^2)}{\sigma_{eff}^{\gamma
p}(Q^2)} \times  \\
&\times \int dx_{p_a} dx_{\gamma_b}
f_{a/p}(x_{p_a}) f_{b/\gamma}(x_{\gamma_b})
d\hat{\sigma}^{2j}_{ab}(x_{p_a},x_{\gamma_b})
\nonumber \\
&\times \int dx_{p_c} dx_{\gamma_d}
f_{c/p}(x_{p_c}) f_{d/\gamma}(x_{\gamma_d})
d\hat{\sigma}^{2j}_{cd}(x_{p_c},x_{\gamma_d})\,.
\nonumber
\end{align}
The sum runs over  active parton flavors and
$d\hat{\sigma}^{2j}$ is the differential partonic cross sections.
Moreover, we tool into account  the unintegrated photon flux $f_{\gamma/e}$ 
\cite{Frixione:1993yw}on  $Q^2$ since 
 $\sigma_{eff}^{\gamma p}$ depends on $Q^2$. 
The distributions $f_{i/A}(x_{A_i})$  are
 the proton ($A=p$) and of the photon ($A=\gamma$) PDFs
for which we use the leading order sets
of Ref. \cite{Pumplin:2002vw} and \cite{Gluck:1991jc}, respectively. We remark 
that the
Dijet cross sections have been  calculated to leading order accuracy by using
{ with \texttt{ALPGEN} \cite{Mangano:2002ea}}. We set both the
factorization and renormalization
scales  to the average transverse  momentum of the jets. As discussed in Ref. 
\cite{Rinaldi:2021vbj}, there are two distinctive
 contributions determined by the
fractional momentum of partons in the photon, $x_\gamma$:
$i)$
 the resolved photon, where 
the photon behaves effectively like an hadron with its own PDF and $ii)$
 the direct one, where the photon interacts as a point-like
particle. One can notice that the first case  populates the whole
$x_\gamma$ range while the latter, at LO, is peaked around at $x_\gamma=1$.
Since the two contributions can mix due to { higher-order} corrections 
\cite{Frixione:1997ks}, here, as in Ref. { \cite{ZEUS:2007njl,H1:2006rre}}, 
we considered specific
kinematic cuts in order to isolate the 
resolved mechanism. In the latter case { ($x_\gamma<0.75$) and a 
direct-enriched
one ($x_\gamma>0.75$).
Therefore, we select, in the DPS cross-section, 
$x_{\gamma,1}+x_{\gamma,2}<0.75$. Furthermore, in order to evaluate the 
possible background, 
in Ref. \cite{Rinaldi:2021vbj}, the four-jet
photoproduction SPS process has been
 evaluated to with \texttt{ALPGEN}
for
$x_\gamma<0.75$, see Table I.
Let us address that
the experimental four-jet photoproduction cross-section, {($\sigma_{exp}$)},
turns to be 135 pb for  $x_\gamma<0.75$  \cite{Chekanov:2007ab} .}
We report in Tab. I  $\sigma_{DPS}$ and
$\sigma_{SPS}$ obtained for three ranges of photon virtualities in HERA
kinematics.
As one can see,
the DPS cross section leads to a sizeable non negligible contribution.
Moreover, 
 as discussed in Ref. \cite{Rinaldi:2021vbj}, 
higher order corrections 
\cite{Klasen:1996it,Klasen:1997br,Badger:2012pf,Bern:2011ep} for both dijet 
photo-production and SPS four-jet emission could provide significant effects, 
see  some qualitative details in Ref. \cite{Rinaldi:2021vbj}.
In this case,  the LO results could 
represent: $i)$ un
upper limit on the SPS background and $ii)$ a lower limit
on the DPS cross section. Therefore, one should 
expect that also high order corrections 
validate the important role of DPS in this process.
For the moment being,  the largest theoretical uncertainty comes 
from the models of 
the proton and photon structures.  
Nevertheless,
a large DPS contribution is predict by all
 models adopted, 
suggesting that jets photoproduction
in $ep$ collisions could represent a golden channel to observe  DPS.

\section{Extraction of the $Q^2$-dependence of $\sigma_{eff}^{\gamma p}$}
Here we discuss the $Q^2$ dependence of the cross-section.
The latter is 
important to extract
information on the proton structure.
We perform such an analysis within the HERA settings presented in the previous
Section.
We  sketch Eq. (\ref{sigmaDPS}) as $ d\sigma_{DPS}(bin)
\sim \int_{bin}dQ^2 g(Q^2)/\sigma_{eff}^{\gamma p}(Q^2)$, where $bin$ stands for
a given interval of  integration over $Q^2$ and the function $g$ encodes the
flux factor, the PDFs and elementary cross sections.
Then we define the ratio $
  R= \sfrac{ d\sigma_{DPS}(bin1)}{d\sigma_{DPS}(bin2)}~.$
{In the case $\sigma_{eff}^{\gamma p}$ were a constant,  the latter
quantity would be: $R \sim \int_{bin1}dQ^2 g(Q^2)/ \int_{bin2}dQ^2
g(Q^2) \sim 2.1$. Therefore,
any discrepancy from this value would
 directly  point to $Q^2$ effects on $\sigma_{eff}^{\gamma p}$ or
possible correlations breaking the pocket formula.
As one can see in the last column of Tab. I, all  models predict
a non trivial dependence of $\sigma_{eff}^{\gamma p}$ on $Q^2$.
{ We also estimated the minimum {
integrated}
luminosity  to experimentally access $Q^2$ effects in
$\sigma_{eff}^{\gamma p}(Q^2)$.
We have converted the cross sections in Tab. I in expected number of
events with a given { integrated} luminosity.
The results are presented in the right panel of Fig. (\ref{Newfig}) where we
pltted two scenarios:
{\it i) }the blue curves
indicate results for the
$\sigma_{eff}^{\gamma p} (Q^2)$; {\it ii)}
the red curve indicate the number of events obtained with a constant, $Q^2$
independent, $\bar \sigma_{eff}^{\gamma p}$ which reproduces the total cross
section
for $Q^2<1$ Ge$\mbox{V}^2$ obtained with $\sigma_{eff}^{\gamma p}(Q^2)$.
Here we make use of the photon proton models which lead to the
the minimal { integrated} luminosity  obtained the two scenarios are
distinguished
 and therefore exposes the $Q^2$-dependence of $\sigma_{eff}^{\gamma p}$ is
$\mathcal{L}=200$ pb$^{-1}$.}
See further details on Ref. \cite{Rinaldi:2021vbj}.

}

%%%%%%%%%%%%%%%%%%% FINE
\section{Conclusions}
\label{conc}

In the present analysis we have calculated the
effective cross sections
for photon induced processes
which are essential ingredients in the predictions
of DPS cross sections in quasi-real photon proton interactions. The latter have
been evaluated by means of 
 electromagnetic and hadronic
models of the photon and proton structures, respectively.
For the four-jet final state in HERA kinematics we found a sizeable DPS
contribution.

In the case the photon virtuality $Q^2$ could be measured, 
we have proven 
that data on $\sigma_{eff}^{\gamma p}$  could be used  to extract new 
information
on the proton structure.
  We set lower limits
on the integrated luminosity needed to observe such a dependence.

\section{Acknowledgements}

This
work was supported, in part by the STRONG-2020 project of the European Unions
Horizon 2020 research and
innovation programme under grant agreement No 824093 and
 by the project “Photon initiated double parton
scattering: illuminating the proton parton structure” on the FRB of the
University of Perugia. The author thank the organizers of the `19th 
International Conference on Hadron Spectroscopy and Structure (HADRON2021)``.

\tabletopline\vspace{2pt}\lilahf{\sc Table I.\ {\rm   \footnotesize Predictions for the LO DPS and SPS cross sections for
four-jet photo-production in three ranges of $Q^2$. In the 
fourth   column, the
ratio between the calculated cross-sections  to the total one is displayed.
In the DPS case, each row corresponds to prediction obtained with a given $pp$
eff ($G_1$,$G_2$,$S$),
and the photon wave function of Refs. \cite{arriola}
 (three upper rows) and Ref.\cite{brodfrank} (three bottom rows). In the last
column the ratio $R$ is shown. }}
\begin{center}
\small{\renewcommand{\arraystretch}{1.2}
\renewcommand{\tabcolsep}{0.5pc}
%\begin{tabular}{|c c|c|c|c|c|}
\begin{tabular}{ccccccc}
\hline
\hline
     \rule{0mm}{0.4cm} %\hhline{~~----}
 & & $  Q^2 \leq 10^{-2}$  & $10^{-2}\leq Q^2 \leq 1$ & $Q^2 \leq 1$  &$
\frac{\displaystyle \sigma  }{\displaystyle \sigma_{exp}}$& $R$ \\
  \rule{0mm}{0.4cm}
  & & [GeV$^2$]  & [GeV$^2$] & [GeV$^2$]  & [\%] & \\  \hline
     \multirow{2}{*}{   }& & \multicolumn{4}{c}{$\sigma_{DPS}$ [pb]}
     \rule{0mm}{0.4cm}
      \\ \hline
\multirow{3}{*}{~w.f. }& \mc{1}{|l|}{G$_1$} & 35.1&
18.6 & 53.7 &40 & 1.89  \\
%\hhline{~-----}
& \mc{1}{|l|}{G$_2$}  & 29.1
   &  15.2 & 44.3 & 33 & 1.91 \\
%\hhline{~-----}
\cite{arriola} & \mc{1}{|l|}{S}  &   26.4                   & 13.7
&
40.1 & 30 & 1.93  \\  \hline
%\hline
\multirow{3}{*}{~w.f. }& \mc{1}{|l|}{G$_1$} & 87.8&
54.3 & 142.1&   101 & 1.62  \\
%\hhline{~-----}
& \mc{1}{|l|}{G$_2$}  &  54.3
   &  33.4 & 87.7 &65  & 1.63 \\
%\hhline{~-----}
\cite{brodfrank}& \mc{1}{|l|}{S}  &   50.5& 31.1 & 81.6 & 60 & 1.62  \\
\hline
\mc{2}{l}{}& \multicolumn{4}{c}{$\sigma_{SPS}$ [pb]}
     \rule{0mm}{0.4cm}
 \\ \hline
%     \rule{0mm}{0.4cm} %\hhline{~~----}
% & & $  Q^2 \leq 10^{-2}$  & $10^{-2}\leq Q^2 \leq 1$ & $Q^2 \leq 1$  &$
%\frac{\displaystyle \sigma_{SPS}}{\displaystyle \sigma_{exp}}$ \\
 % \rule{0mm}{0.4cm}
 % & & [GeV$^2$]  & [GeV$^2$] & [GeV$^2$]  & [\%]  \\  \hline
 \mc{2}{l|}{~~~LO SPS} & 77.5&
36.6 & 114.1 & 86 & 2.12  \\
%\hhline{~-----}
\hline
\hline
\end{tabular}
\label{tab2}
}
\end{center}

\end{multicols}
\medline
%\begin{multicols}{2}

\newpage

\bibliography{glasgow_mpi.bib}

%\end{multicols}

%\bibliography{/home/matteo/Scrivania/Lavoro/latex/REFERENZE/bib_glueball3}

\end{document}